\documentclass[preprint,aps,pra,showpacs,floatfix]{revtex4-1}
\usepackage{graphicx}
\usepackage{times}
\usepackage{nicefrac}
\usepackage{amsmath}
\usepackage{amsfonts}
\usepackage{amssymb}
\usepackage{amsthm}
\usepackage{epsf}
\usepackage{bm}

\usepackage{dcolumn}
\newcolumntype{.}{D{x}{}{-1}}

\newcommand{\be}{\begin{eqnarray}}
\newcommand{\ee}{\end{eqnarray}}

\newcommand{\ket}[1]{|#1\rangle}

\newcommand{\matrixel}[3]{\langle #1 | #2 | #3 \rangle}
\newcommand{\calH}{{\cal H}}
\newcommand{\bcalH}{{\mbox{\boldmath$\cal H$}}}

\newcommand{\balpha}{\bm{\alpha}}

\newcommand{\bfr}{{\mathbf r}}
\newcommand{\bfp}{{\mathbf p}}
\newcommand{\dgrec}{\Delta g_{\mathrm{rec}}}

\newcommand{\dgrece}{\Delta g_{\textrm{non-magn}}}
\newcommand{\dgrecm}{\Delta g_{\textrm{magn}}}
\newcommand{\dgreclo}{\dgrec^{(0)}}
\newcommand{\dgrecz}{\dgrec^{(1)}}
\newcommand{\dgreczz}{\dgrec^{(2+)}}
\newcommand{\dgrecelo}{\dgrece^{(0)}}
\newcommand{\dgrecmlo}{\dgrecm^{(0)}}
\newcommand{\dgrecez}{\dgrece^{(1)}}
\newcommand{\dgrecmz}{\dgrecm^{(1)}}
\newcommand{\dgreck}{\dgrec^{(k)}}
\newcommand{\dgrecek}{\dgrece^{(k)}}
\newcommand{\dgrecmk}{\dgrecm^{(k)}}
\newcommand{\Hmagn}{H_\mathrm{magn}}

\newcommand{\HM}{H_{M}}
\newcommand{\HMmagn}{H_{M}^\mathrm{magn}}
\newcommand{\Hint}{H_{\mathrm{int}}}
\newcommand{\aZ}{\alpha Z}

\begin{document}

\title{Nuclear recoil effect on the \emph{g} factor of middle-$Z$ boronlike ions}
\author{D.~A.~Glazov, A.~V.~Malyshev, V.~M.~Shabaev, and I.~I.~Tupitsyn}

\affiliation{
Department of Physics, St. Petersburg State University, Universitetskaya 7/9, 199034 St.~Petersburg, Russia \\
}

\begin{abstract}
The nuclear recoil correction to the \emph{g} factor of boronlike ions is evaluated within the lowest-order relativistic (Breit) approximation. The interelectronic-interaction effects are taken into account to the first order of the perturbation theory in $1/Z$. Higher orders in $1/Z$ are partly accounted for by means of the effective screening potential. The most accurate up-to-date values of this contribution are presented for the ions in the range $Z=10$--$20$.
\end{abstract}


\maketitle
%
%%%%%%%%%%%%%%%%%%%%%%%%%%%%%%%%%%%%%%%%%%%%%%%%%%%%%%%%%%%%%%%%%%%%%%%%
%
\section{Introduction}
%
%%%%%%%%%%%%%%%%%%%%%%%%%%%%%%%%%%%%%%%%%%%%%%%%%%%%%%%%%%%%%%%%%%%%%%%%
%
The bound-electron \emph{g} factor is the subject of intensive experimental and theoretical investigations during the last 20 years. High-precision measurements for hydrogenlike ions \cite{haefner:00:prl,verdu:04:prl,sturm:11:prl,sturm:13:pra} combined with theoretical calculations (see Ref.~\cite{shabaev:15:jpcrd} and references therein) have lead to the significant improvement of the electron mass value \cite{sturm:14:n,zatorski:17:pra}. Anticipated experiments with few-electron ions are able to deliver an independent determination of the fine-structure constant \cite{shabaev:06:prl,volotka:14:prl-np,yerokhin:16:prl} provided the corresponding progress in theory is achieved. Meanwhile, presently found agreement between theory and experiment manifests the most accurate test of the bound-state QED in the presence of magnetic field \cite{sturm:13:pra,wagner:13:prl,volotka:14:prl,shabaev:15:jpcrd,yerokhin:17:pra-1,yerokhin:17:pra-2}. Even more non-trivial tests are possible in studying the nuclear recoil effect, which demands quantum electrodynamics beyond the Furry picture (i.e., beyond the external field approximation) for its full description. Recent measurement of the \emph{g}-factor isotope shift in lithiumlike calcium \cite{koehler:16:nc} has provided the direct access to the nuclear recoil contribution. Lately we have reevaluated the two-electron part of this contribution and improved the agreement between theory and experiment \cite{shabaev:tbp}. Moreover, we have shown that the non-trivial QED contribution to the nuclear recoil effect can be probed at the few-percent level in the specific difference of the \emph{g} factors of heavy hydrogenlike and lithiumlike ions \cite{malyshev:17:jetpl}.

The \emph{g}-factor measurement performed at the MPIK \cite{soriaorts:07:pra} was the first one sensitive to the QED effects in boronlike systems. The ARTEMIS experiment presently implemented at GSI \cite{lindenfels:13:pra} aims at the precision of $10^{-9}$ for the \emph{g} factors of the ground and the first excited states in boronlike argon. These activities have attracted special attention to the Zeeman splitting in boronlike ions, both to the \emph{g} factor and to the non-linear effects in magnetic field \cite{glazov:13:ps,shchepetnov:15:jpcs,marques:16:pra,agababaev:17:nimb,varentsova:17:nimb}. Present \emph{g}-factor theoretical uncertainty is much larger in this case as compared to lithiumlike ions, in particular, for the ground state of boronlike argon it amounts to $0.7 \times 10^{-6}$ \cite{shchepetnov:15:jpcs}. At the same time, the nuclear recoil effect is much more pronounced for $p$-states ($-9.1 \times 10^{-6}$ for argon) due to the contributions which vanish for $s$-states. So, the \emph{g}-factor investigations for boronlike ions are promising not only for $\alpha$-determination \cite{shabaev:06:prl} but also for testing the QED theory of the nuclear recoil effect.

In this paper we present the most accurate up-to-date relativistic calculations of the nuclear recoil correction to the \emph{g}-factor of boronlike ions. The use of the Dirac wave functions and the corresponding 4-component relativistic operators ensures the result to be complete to orders $m/M$ and $(\aZ)^2 m/M$. The first-order interelectronic-interaction correction is calculated within the Breit approximation. The numerical results are presented for the ions in the range $Z=10$--$20$.

We use the relativistic units ($\hbar=c=1$) throughout the paper.

%
%%%%%%%%%%%%%%%%%%%%%%%%%%%%%%%%%%%%%%%%%%%%%%%%%%%%%%%%%%%%%%%%%%%%%%%%
%
\section{Basic formulae}
%
%%%%%%%%%%%%%%%%%%%%%%%%%%%%%%%%%%%%%%%%%%%%%%%%%%%%%%%%%%%%%%%%%%%%%%%%
%
We consider a boronlike ion in the ground state $(1s)^2\,(2s)^2\,2p_{1/2}$ placed in the constant homogeneous magnetic field $\bcalH$ directed along the $z$ axis. Let $\ket{A}$ be the reference-state many-electron wave function with the energy $E_A$ and the total angular momentum projection $M_J$, calculated to the zeroth-order in $1/Z$. It is the Slater determinant of the Dirac wave functions in the nuclear potential, while $E_A$ is the sum of the corresponding one-electron energies and $M_J$ is the angular momentum projection of the $2p_{1/2}$ state. The magnetic-field interaction is described by the operator
\begin{equation}
  \Hmagn = \mu_0 \bcalH \cdot \sum_{j} \left[ \bfr_j \times \balpha_j \right]
\,,
\end{equation}
where $\mu_0$ is the Bohr magneton, $\balpha$ is the vector of the Dirac matrices.

The non-relativistic operator for the nuclear recoil effect on the bound-electron \emph{g} factor of the first order in the electron-to-nucleus mass ratio $m/M$ was derived by Phillips \cite{phillips:49:pr}. Leading-order relativistic and radiative corrections, as well as the higher orders in $m/M$, were considered in Refs.~\cite{faustov:70:nca,grotch:70:pra,grotch:71:pra,close:71:plb,hegstrom:73:pra,pachucki:08:pra,eides:10:prl}, see also references therein. The fully relativistic theory of this effect in the first order in $m/M$ valid to all orders in $\aZ$ has been developed in Ref.~\cite{shabaev:01:pra}. In the present work, we neglect the non-trivial QED contributions (higher-order part) represented by $\Delta E_\mathrm{H}^{(1,2)}$ from Eqs.~(78) and~(95) of Ref.~\cite{shabaev:01:pra} and consider only the lower-order part represented by $\Delta E_\mathrm{L}^{(1,2)}$ from Eqs.~(77) and~(94) of Ref.~\cite{shabaev:01:pra}. This part is complete to the orders $(\aZ)^0$ and $(\aZ)^2$, while the higher-order part $\Delta E_\mathrm{H}^{(1,2)}$ contains only higher powers of $\aZ$. The lower-order part can be represented by the effective operators, which have to be taken into account in the first order of the perturbation theory. The first one is the nuclear recoil Hamiltonian,
\begin{equation}
\label{eq:HM}
  \HM = \frac{1}{2M}\, \sum_{j,k}
    \left[ \bfp_j\cdot \bfp_k - \frac{\alpha Z}{r_j}
      \left( \balpha_j + \frac{(\balpha_j\cdot\bfr_j)\bfr_j}{r_j^2} \right)
      \cdot \bfp_k \right]
\,,
\end{equation}
which yields the corresponding correction to the binding energy \cite{shabaev:98:pra}. The \emph{g}-factor correction is given by the following second-order perturbation-theory formula:
\begin{equation}
\label{eq:dgrece}
  \dgrecelo = \frac{2}{\mu_0 \calH M_J} \, \sum_{N \neq A}
    \frac{\matrixel{A}{\HM}{N}\matrixel{N}{\Hmagn}{A}}{E_A-E_N}
\,.
\end{equation}
The summation runs over the complete spectrum of the many-electron states $\ket{N}$ constructed as the Slater determinants from the Dirac wave functions, including the negative-energy states.
The second operator \cite{shabaev:tbp},
\begin{equation}
\label{eq:HMmagn}
  \HMmagn = -\mu_0 {\cal H} \frac{m}{M} \sum_{j,k}
    \left\{ [\bfr_j \times \bfp_k] - \frac{\aZ}{2r_k}
      \left[\bfr_j \times
        \left( \balpha_k
          + \frac{(\balpha_k\cdot\bfr_k)\bfr_k}{r_k^2}
        \right)
      \right]
    \right\}
\,,
\end{equation}
arises only in the presence of magnetic field, its contribution to the \emph{g} factor is given by the first-order matrix element,
\begin{equation}
\label{eq:dgrecm}
  \dgrecmlo = \frac{1}{\mu_0 \calH M_J}\, \matrixel{A}{\HMmagn}{A}
\,.
\end{equation}
The first term in the expression (\ref{eq:HMmagn}) for $\HMmagn$ defines the non-relativistic limit of the nuclear recoil effect \cite{phillips:49:pr}. While for $s$-states it yields zero, for $p$-states it gives the dominant contribution for low- and middle-$Z$ ions.

In order to take into account the interelectronic-interaction effects, we consider the first-order correction to $\dgrecelo$ and $\dgrecmlo$ due to the Coulomb-Breit interaction Hamiltonian,
\begin{eqnarray}
\label{eq:Hint}
  \Hint = \alpha \sum_{j<k}
    \left[ \frac{1}{r_{jk}}
           - \frac{1}{2} \left( \frac{ \balpha_j \cdot \balpha_k }{r_{jk}}
             + \frac{ (\balpha_j \cdot \bfr_{jk})
                      (\balpha_k \cdot \bfr_{jk}) }{r_{jk}^3}
           \right)
    \right]
\,.
\end{eqnarray}
The corresponding formula for the magnetic part reads,
\begin{equation}
\label{eq:dgrecmz}
  \dgrecmz = \frac{2}{\mu_0 \calH M_J} \, \sum_{N \neq A}^{+}
    \frac{\matrixel{A}{\HMmagn}{N}\matrixel{N}{\Hint}{A}}{E_A-E_N}
\,,
\end{equation}
where the summation runs over the positive-energy states only, i.e., $\ket{N}$ are constructed as the Slater determinants of the positive-energy one-electron states. The first-order correction $\dgrecez$ to the non-magnetic part is represented by a quite lengthy expression due to the operator permutations and the proper treatment of the negative-energy excitations that shall accompany the operator $\Hmagn$ (discussion of this question can be found, e.g., in Refs.~\cite{glazov:04:pra,tupitsyn:05:pra}).

In addition, an effective screening potential can be introduced into the Dirac equation determining the zeroth-order energies $E_A$, $E_N$ and wave functions $\ket{A}$, $\ket{N}$. In this case, the corresponding counter-term shall be added to $\Hint$ in calculations of $\dgrecez$ and $\dgrecmz$. In this way, the higher-order corrections in $1/Z$ are partly taken into account. We consider the well-known core-Hartree (CH), Kohn-Sham (KS) and local Dirac-Fock (LDF) potentials, for more details see e.g. Refs.~\cite{sapirstein:02:pra,shabaev:05:pra,glazov:06:pla,sapirstein:11:pra,malyshev:17:pra} and references therein. 

The recursive approach to evaluate the interelectronic-interaction corrections within the Breit approximation to all orders in $1/Z$ has been developed in Refs.~\cite{glazov:17:nimb,shabaev:tbp}. It has been employed to evaluate these corrections to the nuclear recoil effect on the \emph{g}-factor of lithiumlike ions~\cite{shabaev:tbp,malyshev:17:jetpl}. This method can be applied also to the present case of boronlike ions, which will be the subject of our subsequent research.

%
%%%%%%%%%%%%%%%%%%%%%%%%%%%%%%%%%%%%%%%%%%%%%%%%%%%%%%%%%%%%%%%%%%%%%%%%
%
\section{Results and discussion}
%
%%%%%%%%%%%%%%%%%%%%%%%%%%%%%%%%%%%%%%%%%%%%%%%%%%%%%%%%%%%%%%%%%%%%%%%%
%
The numerical calculations of the leading-order nuclear-recoil terms, $\dgrecelo$ and $\dgrecmlo$, and the first-order corrections due to the interelectronic interaction, $\dgrecez$ and $\dgrecmz$, are performed employing the finite-basis-set method. The spectrum of the Dirac-equation solutions for the Coulomb nuclear potential or for one of the effective potentials is found within the dual-kinetic-balance (DKB) approach \cite{shabaev:04:prl}. The many-electron wave functions $\ket{N}$ including the reference state $\ket{A}=\ket{(1s)^2\,(2s)^2\,2p_{1/2}}$ are constructed as the Slater determinants from these one-electron wave functions. In Table~\ref{tab:g-rec-ic} the individual contributions to the nuclear recoil correction in boronlike argon are presented. The zeroth- and first-order coefficients $A(\aZ)$ and $B(\aZ)$ are defined as
\begin{align}
\label{eq:A}
  &\dgreclo = \frac{m}{M} A(\aZ)
\,,\\
% \hspace{1cm}
\label{eq:B}
  &\dgrecz = \frac{m}{M} \frac{B(\aZ)}{Z}
\,,\\
  &\dgreck = \dgrecek + \dgrecmk
\,.
\end{align}
We note, however, that $A$ and $B$ defined in this way represent the $1/Z$-expansion only in the case of the Coulomb potential. For the screening potentials, these coefficients incorporate partly the higher orders in $1/Z$ and shall be written as $A(\aZ,Z)$ and $B(\aZ,Z)$. The contribution of the magnetic-recoil operator $\HMmagn$ is divided into the non-relativistic (first term in Eq.~(\ref{eq:HMmagn})) and relativistic (second term in Eq.~(\ref{eq:HMmagn})) parts. Moreover, it is divided into the one-electron ($j=k$) and two-electron ($j\neq k$) parts, so there are four in total: ``magn 1-el'', ``magn 2-el'', ``magn 1-el-r'', and ``magn 2-el-r''. The ``non-magn'' contribution of the operator $\HM$ is completely relativistic and is given undivided in Table~\ref{tab:g-rec-ic}.

The results for even values of $Z$ in the range $Z=10$--$20$ are given in Table~\ref{tab:g-rec-pot} in terms of the coefficients $A$ and $B/Z$ for the Coulomb and three different screening potentials. As the final result, we take the value for the LDF potential. One can see that the difference between the values for different potentials gets several times smaller when the first-order correction $B/Z$ is taken into account. On the other hand, the screening effect for the zeroth-order term $A$ accounts for about $50\%$ only of the total interelectronic-interaction effect obtained. The reason for this is the structure of the two-electron ``ladder'' terms in Eq.~(\ref{eq:dgrecmz}) with the two-electron magnetic-recoil operator, which can not be approximated by the terms with the screening potential in place of $\Hint$. This is in contrast, e.g., to the leading-order (non-recoil) interelectronic-interaction correction to the \emph{g} factor, where the screening potential accounts for the dominant part of the interelectronic-interaction effect, and the spread of the results for different potentials can serve as an estimation of the higher-order terms \cite{volotka:14:prl}. So, in the present case such an estimation of the unknown $1/Z^{2+}$ contribution will not work most probably. For this reason, we estimate it from the ratio of $\dgrecz$ to $\dgreclo$ for the Coulomb potential: $\dgreczz \sim \dgrecz \cdot (\dgrecz/\dgreclo)$.

Another source of the uncertainty is the non-trivial QED part of the nuclear recoil correction, which may give terms of the order $(\aZ)^3$ and higher (for $s$-states only $(\aZ)^5$ and higher) \cite{shabaev:01:pra}. First evaluation of this contribution for the $1s$ state complete to all orders in $\aZ$ was performed in Ref.~\cite{shabaev:02:prl}. Calculations for the $2s$ state have been done in Ref.~\cite{koehler:16:nc} for lithiumlike calcium and in Refs.~\cite{shabaev:tbp,malyshev:17:jetpl} for the ions in the range $Z=3$--$92$. Estimated as $(\aZ)^3 \dgreclo$ it appears to be about $0.002$ for $Z=20$ in terms of the $A$ and $B/Z$ coefficients, that is much smaller than the uncertainty due to the higher-order interelectronic interaction ($\dgreczz$). It should be noted that for boronlike ions there exists also a two-electron contribution in the zeroth order in $1/Z$ which is beyond the Breit approximation. It was calculated in Ref.~\cite{shchepetnov:15:jpcs} for boronlike argon. Due to its smallness, however, we neglect it in the present work. We also neglect the so-called radiative corrections ($\sim \alpha\, m/M$) and the contributions of the higher orders in $m/M$.

Finally, in Table~\ref{tab:g-rec} the nuclear recoil correction $\dgrec=\dgreclo+\dgrecz$ to the \emph{g} factor of several boronlike ions in the range $Z=10$--$20$ is presented. The values of the $A$ and $B$ coefficients obtained with the LDF potential are used, the estimation of the uncertainty is described above. The result for boronlike argon slightly differs from the one presented in Ref.~\cite{shchepetnov:15:jpcs} due to the relativistic corrections to the $1/Z$ term.

% motivated by the corresponding measurement sensitive to the recoil correction on the level of $10\%$.

%
%%%%%%%%%%%%%%%%%%%%%%%%%%%%%%%%%%%%%%%%%%%%%%%%%%%%%%%%%%%%%%%%%%%%%%%%
%
\section{Conclusion}
%
%%%%%%%%%%%%%%%%%%%%%%%%%%%%%%%%%%%%%%%%%%%%%%%%%%%%%%%%%%%%%%%%%%%%%%%%
%
The nuclear recoil effect on the \emph{g} factor of middle-$Z$ boronlike ions is evaluated in the first order in $m/M$ and in the zeroth and first orders in $1/Z$. The leading relativistic corrections of the order $(\aZ)^2$ are taken into account employing the relativistic nuclear recoil operators derived in Refs.~\cite{shabaev:98:pra,shabaev:01:pra}. The interelectronic-interaction correction of the first order in $1/Z$ is evaluated within the perturbation theory. The higher-order contributions in $1/Z$, which currently determine the theoretical uncertainty, are partly included by means of the effective screening potential.

%
%%%%%%%%%%%%%%%%%%%%%%%%%%%%%%%%%%%%%%%%%%%%%%%%%%%%%%%%%%%%%%%%%%%%%%%%
%
\section{Acknowledgments}
%
%%%%%%%%%%%%%%%%%%%%%%%%%%%%%%%%%%%%%%%%%%%%%%%%%%%%%%%%%%%%%%%%%%%%%%%%
%
This work was supported by the Russian Science Foundation (Grant No. 17-12-01097).
%
%%%%%%%%%%%%%%%%%%%%%%%%%%%%%%%%%%%%%%%%%%%%%%%%%%%%%%%%%%%%%%%%%%%%%%%%
%  References
%%%%%%%%%%%%%%%%%%%%%%%%%%%%%%%%%%%%%%%%%%%%%%%%%%%%%%%%%%%%%%%%%%%%%%%%
%

%
%%%%%%%%%%%%%%%%%%%%%%%%%%%%%%%%%%%%%%%%%%%%%%%%%%%%%%%%%%%%%%%%%%%%%%%%
%
%%%%%%%%%%%%%%%%%%%%%%%%%%%%%%%%%%%%%%%%%%%%%%%%%%%%%%%%%%%%%%%%%%%%%%%%
%
%%%%%%%%%%%%%%%%%%%%%%%%%%%%%%%%%%%%%%%%%%%%%%%%%%%%%%%%%%%%%%%%%%%%%%%%
% Results
%%%%%%%%%%%%%%%%%%%%%%%%%%%%%%%%%%%%%%%%%%%%%%%%%%%%%%%%%%%%%%%%%%%%%%%%
%
%
\setlength{\tabcolsep}{12pt}
%
%
%%%%%%%%%%%%%%%%%%%%%%%%%%%%%%%%%%%%%%%%%%%%%%%%%%%%%%%%%%%%%%%%%%%%%%%%
%  Individual contributions
%%%%%%%%%%%%%%%%%%%%%%%%%%%%%%%%%%%%%%%%%%%%%%%%%%%%%%%%%%%%%%%%%%%%%%%%
%
%
\begin{table}
\caption{Individual contributions to the zeroth- and first-order terms in $1/Z$ of the nuclear recoil correction to the ground-state \emph{g} factor of boronlike argon in the Coulomb, core-Hartree (CH), Kohn-Sham (KS), and local Dirac-Fock (LDF) potentials. Terms $A$ and $B/Z$ are defined by Eqs.~(\ref{eq:A}) and~(\ref{eq:B}).
\label{tab:g-rec-ic}}
\vspace{0.5cm}
\begin{tabular}{clr@{}lr@{}lr@{}lr@{}l}
\hline
Term
& Contribution
& \multicolumn{2}{c}{Coulomb}
& \multicolumn{2}{c}{CH}
& \multicolumn{2}{c}{KS}
& \multicolumn{2}{c}{LDF}
\\
\hline
$A$         &
magn 1-el   &   $-$1.&331 888
            &   $-$1.&332 242
            &   $-$1.&332 252
            &   $-$1.&332 253
\\
            &
magn 2-el   &      0.&551 859
            &      0.&600 391
            &      0.&608 808
            &      0.&607 783
\\
            &
magn 1-el-r &      0.&002 888
            &      0.&002 508
            &      0.&002 496
            &      0.&002 495
\\
            &
magn 2-el-r &   $-$0.&002 607
            &   $-$0.&002 190
            &   $-$0.&002 161
            &   $-$0.&002 185
\\
            &
non-magn    &      0.&003 313
            &      0.&002 877
            &      0.&002 937
            &      0.&002 859
\\
\hline
$B/Z$       &
magn 1-el   &   $-$0.&000 347
            &      0.&000 021
            &      0.&000 031
            &      0.&000 031
\\
            &
magn 2-el   &      0.&103 704
            &      0.&067 337
            &      0.&057 242
            &      0.&059 204
\\
            &
magn 1-el-r &   $-$0.&000 326
            &      0.&000 046
            &      0.&000 059
            &      0.&000 058
\\
            &
magn 2-el-r &      0.&000 147
            &   $-$0.&000 286
            &   $-$0.&000 313
            &   $-$0.&000 289
\\
            &
non-magn    &   $-$0.&000 424
            &   $-$0.&000 019
            &   $-$0.&000 099
            &   $-$0.&000 003
\\
\hline
\end{tabular}
\end{table}
%
%
%%%%%%%%%%%%%%%%%%%%%%%%%%%%%%%%%%%%%%%%%%%%%%%%%%%%%%%%%%%%%%%%%%%%%%%%
%  A and B/Z for Z=10-20
%%%%%%%%%%%%%%%%%%%%%%%%%%%%%%%%%%%%%%%%%%%%%%%%%%%%%%%%%%%%%%%%%%%%%%%%
%
%
\begin{table}
\caption{Results for the zeroth- and first-order terms in $1/Z$ of the nuclear recoil correction to the ground-state \emph{g} factor of boronlike ions in the Coulomb, core-Hartree (CH), Kohn-Sham (KS), and local Dirac-Fock (LDF) potentials. Terms $A$ and $B/Z$ are defined by Eqs.~(\ref{eq:A}) and~(\ref{eq:B}).
\label{tab:g-rec-pot}}
\vspace{0.5cm}
\begin{tabular}{ccr@{}lr@{}lr@{}lr@{}l}
\hline
$Z$
& Term
& \multicolumn{2}{c}{Coulomb}
& \multicolumn{2}{c}{CH}
& \multicolumn{2}{c}{KS}
& \multicolumn{2}{c}{LDF}
\\
\hline
10          &
$A$         &   $-$0.&777 808
            &   $-$0.&672 183
            &   $-$0.&651 781
            &   $-$0.&656 351
\\
            &
$B/Z$       &      0.&185 684
            &      0.&121 905
            &      0.&093 367
            &      0.&103 124
\\
            &
$A+B/Z$     &   $-$0.&592 125
            &   $-$0.&550 278
            &   $-$0.&558 414
            &   $-$0.&553 227
\\
\hline
12          &
$A$         &   $-$0.&777 543
            &   $-$0.&696 297
            &   $-$0.&681 023
            &   $-$0.&683 882
\\
            &
$B/Z$       &      0.&154 621
            &      0.&101 798
            &      0.&081 679
            &      0.&087 482
\\
            &
$A+B/Z$     &   $-$0.&622 922
            &   $-$0.&594 499
            &   $-$0.&599 344
            &   $-$0.&596 400
\\
\hline
14          &
$A$         &   $-$0.&777 227
            &   $-$0.&711 306
            &   $-$0.&699 194
            &   $-$0.&701 166
\\
            &
$B/Z$       &      0.&132 414
            &      0.&087 011
            &      0.&071 696
            &      0.&075 548
\\
            &
$A+B/Z$     &   $-$0.&644 813
            &   $-$0.&624 295
            &   $-$0.&627 498
            &   $-$0.&625 618
\\
\hline
16          &
$A$         &   $-$0.&776 858
            &   $-$0.&721 440
            &   $-$0.&711 447
            &   $-$0.&712 902
\\
            &
$B/Z$       &      0.&115 742
            &      0.&075 821
            &      0.&063 556
            &      0.&066 313
\\
            &
$A+B/Z$     &   $-$0.&661 116
            &   $-$0.&645 619
            &   $-$0.&647 891
            &   $-$0.&646 589
\\
\hline
18          &
$A$         &   $-$0.&776 436
            &   $-$0.&728 656
            &   $-$0.&720 172
            &   $-$0.&721 301
\\
            &
$B/Z$       &      0.&102 759
            &      0.&067 099
            &      0.&056 919
            &      0.&059 002
\\
            &
$A+B/Z$     &   $-$0.&673 677
            &   $-$0.&661 557
            &   $-$0.&663 253
            &   $-$0.&662 299
\\
\hline
20          &
$A$         &   $-$0.&775 957
            &   $-$0.&733 979
            &   $-$0.&726 622
            &   $-$0.&727 531
\\
            &
$B/Z$       &      0.&092 359
            &      0.&060 121
            &      0.&051 451
            &      0.&053 088
\\
            &
$A+B/Z$     &   $-$0.&683 598
            &   $-$0.&673 858
            &   $-$0.&675 171
            &   $-$0.&674 443
\\
\hline
\end{tabular}
\end{table}
%
%
%%%%%%%%%%%%%%%%%%%%%%%%%%%%%%%%%%%%%%%%%%%%%%%%%%%%%%%%%%%%%%%%%%%%%%%%
%  Nuclear recoil correction for Z=10-20
%%%%%%%%%%%%%%%%%%%%%%%%%%%%%%%%%%%%%%%%%%%%%%%%%%%%%%%%%%%%%%%%%%%%%%%%
%
%
\begin{table}
\caption{Nuclear recoil correction to the ground-state \emph{g} factor of middle-$Z$ boronlike ions evaluated as $\dgrec=\dgreclo+\dgrecz$. The uncertainty ascribed to $A+B/Z$ ($A$ and $B$ are defined by Eqs.~(\ref{eq:A}) and~(\ref{eq:B})) is due to the uncalculated higher orders in $1/Z$.
\label{tab:g-rec}}
\vspace{0.5cm}
\begin{tabular}{cr@{}lr@{}lr@{}l}
\hline
Ion
& \multicolumn{2}{c}{$m/M \times 10^6$}
& \multicolumn{2}{c}{$A+B/Z$}
& \multicolumn{2}{c}{$\dgrec \times 10^6$}
\\
\hline
${}^{20}_{10}$Ne$^{5+}$
         &     27.&447
         &   $-$0.&553 (44)
         &  $-$15.&2 (12)
\\
${}^{24}_{12}$Mg$^{7+}$
         &     22.&878
         &   $-$0.&596 (31)
         &  $-$13.&64 (70)
\\
${}^{28}_{14}$Si$^{9+}$
         &     19.&614
         &   $-$0.&626 (23)
         &  $-$12.&27 (44)
\\
${}^{32}_{16}$S$^{11+}$
         &     17.&163
         &   $-$0.&647 (17)
         &  $-$11.&10 (30)
\\
${}^{40}_{18}$Ar$^{13+}$
         &     13.&731
         &   $-$0.&662 (14)
         &   $-$9.&09 (19)
\\
${}^{40}_{20}$Ca$^{15+}$
         &     13.&731
         &   $-$0.&674 (11)
         &   $-$9.&26 (15)
\\
${}^{48}_{20}$Ca$^{15+}$
         &     11.&443
         &   $-$0.&674 (11)
         &   $-$7.&72 (13)
\\
\hline
\end{tabular}
\end{table}
%
%
% \begin{table}
% \caption{Individual contributions to the ground-state \emph{g} factor of boron-like argon.
% \label{tab:g-th-Ar}}
% \vspace{0.5cm}
% \begin{tabular}{llr@{}l}
% %
% %
% \hline
% %
% Dirac value                           &&    0.&663 775 447  \\
% %
% Finite nuclear size                   &&    0.&000 000 000  \\
% %
% One-photon exchange  & $\sim 1/Z$      &    0.&000 657 525  \\
% %
% Many-photon exchange & $\sim 1/Z^{2+}$ & $-$0.&000 007 5 (4)\\
% %
% One-loop QED     & $\sim \alpha$       & $-$0.&000 769 9 (5)\\
% %
% Higher-order QED & $\sim \alpha^{2+}$  &    0.&000 001 2 (1)\\
% %
% Nuclear recoil                        && $-$0.&000 009 1 (2)\\
% %
% \hline
% %
% Total                                 &&    0.&663 647 7 (7)\\
% %
% \hline
% %
% \end{tabular}
% \end{table}
%
\end{document}